\documentclass[acus]{epac98}  
\usepackage{epsfig}
\newcommand{\bit}{\begin{Itemize}}
\newcommand{\eit}{\end{Itemize}}
\begin{document}
\title{Muon Collider Overview: Progress and Future Plans}

\author{R.~Palmer, A.~Sessler, A.~Tollestrup and \underline{J.~Gallardo} for the Muon Collider Collaboration,\\ BNL, New York, USA}
\maketitle

\begin{abstract} 
 Besides
continued work on the parameters of a 3-4 and 0.5~TeV center of mass (CoM)
collider, many studies are now concentrating on a machine near 100~GeV
(CoM) that could be a factory for the $s$-channel production of Higgs
particles.  We mention the research on the various components in such
muon colliders, starting from the proton accelerator needed to
generate pions from a heavy-$Z$ target and proceeding through the phase
rotation and decay ($\pi \rightarrow \mu\,\nu_{\mu} $) channel, muon cooling,
acceleration, storage in a collider ring and the collider
detector. We also mention theoretical and experimental R \& D
plans for the next several years that should lead to a better 
understanding of the design and feasibility issues for all of the
components. This note is a summary of a report\cite{ref1} updating the progress on the R \& D since the Feasibility Study of Muon Colliders presented at the
Workshop Snowmass'96.\cite{ref1a}

\end{abstract}

\section{INTRODUCTION}
\begin{figure}[hb!]
\centering
\epsfig{file=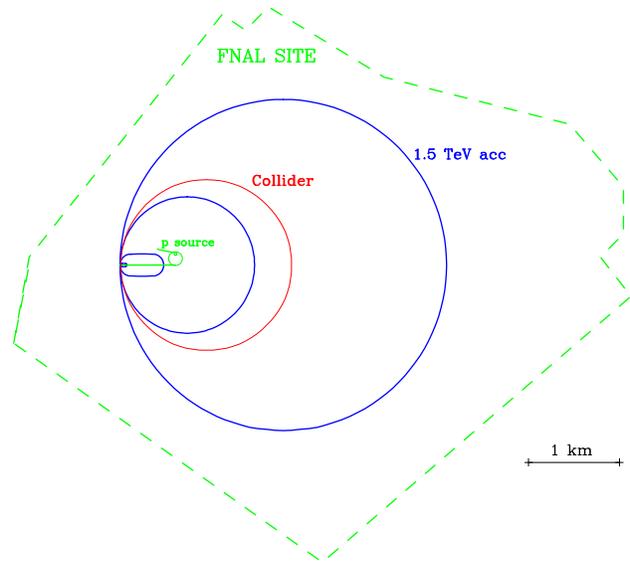, width=82.5mm}
\caption{Plan of a 3~TeV Muon Collider shown on the FNAL site as an example.}
\label{fg1}
\end{figure}

\begin{figure}[hb!]
\centering
\epsfig{file=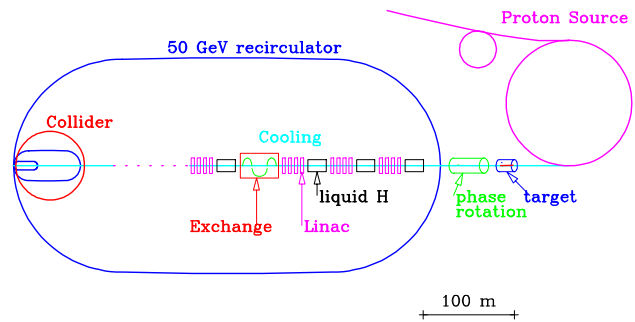, width=82.5mm}
\caption{Plan of a 100~GeV CoM Muon Collider.}
\label{fg2}
\end{figure}

Unlike protons, muons are point like but, unlike electrons, they emit
relatively little synchrotron radiation and therefore, can be accelerated and collided in
rings. As a result, a muon collider with a given energy reach could be smaller
than either a proton or electron  machine. A 3 TeV muon collider (with
effective energy comparable with that of an SSC) would fit on existing sites, such as
BNL or FNAL (see Figs.\ref{fg1}, \ref{fg2}). Another advantage resulting from the low synchrotron radiation is
the lack of beamstrahlung and the possibility of very small collision energy
spreads. A beam energy of $\Delta$E/E of  0.003 \% (equivalent to a CoM spread of $\Delta$E/E of  0.002 \%) is considered feasible for a 100~GeV machine; and
it has been shown that by observing spin precession, the absolute energy could
be determined to a small fraction of this width. These features become 
important in conjuction with the large s-channel Higgs production ($\mu^+\mu^-\rightarrow h$, 
43000 times larger than for $e^+e^-\rightarrow h$), allowing precision
measurements of the Higgs mass, width and branching ratios.

Such machines are clearly desirable. The questions are:
\bit 
\item {\bf whether they can be built and physics done with them}
\item {\bf what will they cost}.
\eit
 Much progress has
been made in addressing the first question and the answer, so far, appears to
be positive. It is too early yet to address the second. 
 We have studied
machines with center of mass energies of 100~GeV, 400~GeV and 3~TeV, defined
parameters and simulated many of their components (see Tb.\ref{sum}). Most work has been done on
the 100~GeV ``First Muon Collider", the exact energy taken to be representative
of the actual mass of a Higgs particle.
\section{COMPONENTS}
\subsubsection{Proton  Driver}
The specification of the proton driver for the three machines is assumed the
same: $10^{14}$  protons/pulse at an energy above 16~GeV and 1-2~ns rms bunch
lengths. There have been three studies of how to achieve them. The most
conservative, at 30~GeV, is a generic design. Upgrades of the FNAL (at 16~GeV)
and BNL (at 24~GeV) accelerators have also been studied. Despite the very short bunch
requirement, each study has concluded that the specification is attainable.
Experiments have been done and are planned to confirm some aspects of these designs.\cite{refproton}
\subsubsection{Muon Production}
Pion production has been taken from the best models available, but an
experiment (BNL-E910) that has taken data, and is being analyzed, will refine these
models.\cite{refe910} The assumed 20~T capture solenoid appears to be well within current
technology (a coil with the specified field and aperture is now nearing
completion at the National High Magnetic Field Laboratory, Florida State University). Capture, decay and phase rotation have been simulated, and
have achieved the specified production of 0.3 muons per initial proton. The
most serious remaining questions for this part of the machine are:
\begin{enumerate}
\item {\bf The nature and material of the target:}
The baseline assumption is that a liquid metal jet will be used, but the
effects of shock heating by the beam, and of the eddy currents induced in the
liquid as it enters the solenoid, are not yet fully understood.  \item {\bf The
maximum RF field in the phase rotation.} For the short pulses used, the current
assumptions would be reasonably conservative under normal operating conditions,
but the effects of the massive radiation from the nearby target are not known.
\end{enumerate}
Both these questions can be answered in a target experiment planned to be
performed within the next two years at AGS.\cite{refkirkmc}

\subsubsection{Cooling}
The required ionization cooling is the most difficult and least understood
element in any of the muon colliders studied.
 Ionization cooling is a phenomenon that occurs whenever there is energy
loss in a strong focusing environment. Such an environment has existed, for
instance, in the iron toroid muon calorimeters of several neutrino
experiments, and a Monte Carlo simulation has shown\cite{ref41} that cooling must
have occured there. But achieving the
nearly $10^6$ reduction required is a challenge. Cooling over a wide range has
been simulated using lithium lenses and ideal (linear matrix) matching and
acceleration; and examples of limited sections of solenoid lattices with
realistic accelerating fields have now been simulated. But the specification
and simulation of a complete system has not yet been done. Much theoretical
work remains: space  charge and wake fields must be included; lattices at the
start and end of the cooling sequences must be designed; lattices including
liquid lithium lenses must be designed and studied, and the sections must be
matched together and simulated as a full sequence. The tools for this work
are nearly ready, and this project should be completed within two years.\cite{reficool}

Technically, one of the most challenging aspects of the cooling system appear
to be: 
\bit
\item {\bf High gradient RF} (e.g. 36~MV/m at 805~MHz) operating in strong
(5-10~T) magnetic field, with beryllium foils between the cavities. 
\eit
An experiment is planned that will test such a cavity, in the required fields,
in about two years time. On an approximately six year time scale, a ``Cooling
Test Facility" is being proposed that could test ten meter lengths of different
cooling systems.\cite{refsgeer} If they are required, there is the need to develop:
\bit
\item {\bf Lithium Lenses:} (e.g.\ 2~cm diameter, 70~cm long, liquid lithium
lenses with 10~T surface fields and a repetition rate of 15~Hz).
\eit
They may not be needed for the low energy ``First Muon Collider", which would
ease the urgency of this rather long term R \& D. Meanwhile a short lithium lens
is under construction at BINP (Novosibirsk, Russia).
\begin{table*}[thb]
\centering  
\caption{Baseline parameters for high and low energy~muon~colliders.}
\label{sum}
\begin{tabular}{|l|c|c|c|c|c|}
\hline\hline
\vspace{1.mm}
 CoM energy (TeV)   & 3 & 0.4 & \multicolumn{3}{c|}{0.1 }  \\
\cline{4-6}
\hline
$p$ energy (GeV)        &  16  & 16 & \multicolumn{3}{c|}{16}\\
$p$'s/bunch  &  $2.5\times 10^{13}$  & $2.5\times 10^{13}$  & \multicolumn{3}{c|}{$5\times 10^{13}$  }  \\  
Bunches/fill & 4 & 4 & \multicolumn{3}{c|}{2 }  \\
Rep.~rate (Hz)     &  15 & 15 & \multicolumn{3}{c|}{15 }  \\
$p$ power (MW)         &  4   & 4 & \multicolumn{3}{c|}{4}  \\ 
$\mu$/bunch & $2\times 10^{12}$ & $2\times 10^{12}$ &   \multicolumn{3}{c|}{$4\times 10^{12}$ }  \\
 $\mu$ power (MW)     &  28 & 4 & \multicolumn{3}{c|}{ 1 }  \\
 Wall power (MW)    &   204 & 120  & \multicolumn{3}{c|}{81 }  \\
Collider circum. (m) &  6000 & 1000 & \multicolumn{3}{c|}{300 }  \\
Depth (m) &  500 & 100 & \multicolumn{3}{c|}{10 }  \\
\cline{4-6}
Rms ${\Delta p\over p}$ (\%) & 0.16 & 0.14 & 0.12 & 0.01&0.003\\
6d $\epsilon_6$ ($(\pi \textrm{m})^3$)&$1.7\times 10^{-10}$&$1.7\times 10^{-10}$&$1.7\times 10^{-10}$&$1.7\times 10^{-10}$&$1.7\times 10^{-10}$\\
Rms $\epsilon_n$ ($\pi$ mm mrad)&  50 & 50 & 85 & 195 & 280\\
$\beta^*$ (cm) & 0.3 & 2.3 & 4 &  9 & 13\\
$\sigma_z$ (cm) & 0.3 & 2.3 & 4 &  9 & 13 \\
$\sigma_r$ spot ($\mu m$)     & 3.2 & 24 & 82 & 187 & 270\\
Tune shift    &0.043 &0.043 & 0.05 &0.02 & 0.015\\
 Luminosity ($cm^{-2}s^{-1}$)& $5\times 10^{34}$ & $10^{33}$ &
$1.2\times 10^{32}$ & $2\times 10^{31}$& $10^{31}$ \\
CoM ${\Delta E\over E}$ & $8\times 10^{-4}$ & $8\times 10^{-4}$ & $8\times 10^{-4}$ & $7\times 10^{-5}$ & $2\times 10^{-5}$ \\ 
\hline
Higgs/year &  & & $1.6\times 10^3$ & $4\times 10^3$ & $4\times 10^3$ \\
\hline
\end{tabular}
\end{table*}
\subsubsection{Acceleration}

The acceleration systems are probably the least controversial, although
possibly the most expensive, part of a muon collider. Preliminary parameters
have been specified for acceleration sequences for a 100~GeV and 3~TeV
machines, but they need refinement. In the low energy case a linac is followed
by three recirculating accelerators. In the high energy accelerator, the
recirculating accelerators are followed by three fast ramping synchrotrons
employing alternating pulsed and superconducting magnets. The parameters do
not appear to be extreme, and it does not appear as if serious problems are
likely. 

\subsubsection{Collider}

The collider lattices are challenging because of their required very low
intersection betas, high single bunch intensities, and short bunch lengths (see Tb.\ref{sum});
however, the fact that all muons will decay after about 1000 turns means that
slowly developing instability   are not a problem. Feasibility lattices have
been generated for a 4~TeV case, and more detailed designs for 100~GeV machines
studied. In the latter case, but still without errors, $5 \sigma$~acceptances in
both transverse and longitudinal phase space have been achieved in tracking
studies. Beam scraping schemes have been designed for both the low energy
(collimators) and high energy (septum extractors) cases.

Bunch length and longitudinal stability problems are avoided if the rings, as
specified, are sufficiently isochronous, but some rf is needed to remove the
impedance generated momentum spread. Transverse instabilities (beam breakup)
should be controlled by rf BNS damping.

The heating of collider ring superconducting magnets by electrons from muon
decay can be controlled by thick tungsten shields, and this technique also
shields the space surrounding the magnets from the induced radioactivity on the
inside of the shield wall. A conceptual design of magnets for the low energy
machine has been defined.

Although much work is yet to be done (inclusion of errors, higher order
correction, magnet design, rf design, etc), the collider ring do not appear
likely to present serious problems.

\subsubsection{Neutrino Radiation and Detector Background}

Neutrino radiation, which naturally rises as the cube of the energy, is not 
serious for machines with center of mass energies below about 1.5~TeV. It is
thus not significant for the First Muon Collider; but above 2~TeV, it
sets a constraint on the muon current and makes it harder to achieve desired
luminosities. However, advances in cooling, and correction of tune shifts may
still allow a machine at 10~TeV with substantial luminosity ($>\ 10^{35}\
\textrm{cm}^{-2}\textrm{s}^{-1}$).

Background in the detector was, at first, expected to be a very serious
problems. But after much work, shielding systems have evolved that limit most
charged hadron, electron, gamma and neutron background to levels that are
expected to be acceptable. Muon background, in the higher energy machines, is a
special problem that can cause serious fluctuations in calorimeter
measurements. It has been shown that fast timing and segmentation can help
suppress this background, and preliminary studies of its effects on a physics
experiment are encouraging. The studies are ongoing.\cite{refstumer}
\section{SUMMARY}
Much progress has been made since Snowmass, but much still needs to be done. A
time scale of two years should allow completion of simulation studies, and the
experimental testing of crucial technical challenges. Prototype
construction and testing will be required for another 4-6~years. The
construction of a ``First Muon Collider" by about 2010 does seem to be possible.
\section{ACKNOWLEDGMENTS}
This research was supported by the U.S. Department of Energy under Contracts No.
DE-ACO2-98CH10886, DE-AC02-76CH03000 and DE-AC03-76SF00098.


\begin{thebibliography}{9}
\bibitem{ref1} {\em Status of the Muon Collider Research and Development and Future Plans}, in preparation.
\bibitem{ref1a} R.~B.~Palmer, A.~Sessler and A.~Tollestrup, {\em Proceedings of the 1996 DPF/DPB Summer Study on High-Energy Physics} (Stanford Linear Accelerator Center, Menlo Park, CA, 1997)
\bibitem{refproton}C.~Ankenbrandt et~al., Bunching Near Transition in the AGS, Fermilab
  Pub-98-006, submitted to Phys. Rev. D.

\bibitem{refe910}
Experiment E-910 at~BNL-AGS, available at the URL \hfill\break http://www.nevis.columbia.edu/heavyion/e910/,
  1997.

\bibitem{refkirkmc} K.~McDonald, available at the URL \hfill\break http://www.hep.princeton.edu/mumu/
\bibitem{ref41} B.~King, available at the URL \hfill\break http://pubweb.bnl.gov/people/bking/ 
\bibitem{reficool}R.~Fernow, \textit{ICOOL}, Fortran program to simulate muon ionization cooling.
\bibitem{refsgeer}C.~Ankenbrandt et~al., \textit{Ionization Cooling R\&D Program for a High Luminosity Muon Collider} (MUCOOL Proposal), FNAL-P904, April 1998; available at the URL \hfill\break
http://www.fnal.gov/projects/muon$_{}$collider/
\bibitem{refstumer}I. Stumer et al., \textit{Study of Detector Backgrounds in a $\mu^+\mu^-$ Collider }, Proceedings of the 1996 DPF/DPB Summer Study on High-Energy Physics, (Stanford Linear Accelerator Center, Menlo Park, CA, 1997)
\end{thebibliography}
\end{document}